\documentclass[aps,prl,twocolumn,superscriptaddress,preprintnumbers,longbibliography,nofootinbib]{revtex4-1}

\usepackage{amssymb}
\usepackage{graphicx}
\usepackage{amsmath}
\usepackage{hyperref}
\usepackage{xspace}

\DeclareRobustCommand{\Fig}[1]{Fig.~\ref{#1}}

\DeclareRobustCommand{\Eq}[1]{eq.\,(\ref{#1})}

\DeclareRobustCommand{\Ref}[1]{ref.\,\cite{#1}}

\usepackage{color}
\definecolor{darkblue}{rgb}{0,0,0.5}
\definecolor{darkred}{rgb}{0.5,0,0}
\definecolor{darkgreen}{rgb}{0,0.5,0}

\newcommand{\be}{\begin{equation}}
\newcommand{\ee}{\end{equation}}

\begin{document}

\preprint{}

\title{Theory Predictions for the Pull Angle}

\author{Andrew J.~Larkoski}
\email{larkoski@reed.edu}

\affiliation{Physics Department, Reed College, Portland, OR 97202, USA}

\author{Simone Marzani}
\email{simone.marzani@ge.infn.it}
\affiliation{Dipartimento di Fisica, Universit\`a di Genova and INFN, Sezione di Genova,\\ Via Dodecaneso 33, 16146, Italy}

\author{Chang Wu}
\email{chang.wu@ge.infn.it}
\affiliation{Dipartimento di Fisica, Universit\`a di Genova and INFN, Sezione di Genova,\\ Via Dodecaneso 33, 16146, Italy}

\begin{abstract}
Pull is a jet observable that is sensitive to color flow between dipoles.  It has seen wide use for discrimination of particles with similar decay topologies but carrying different color representations and has been measured on $W$ bosons from top quark decays by the D$\emptyset$ and ATLAS experiments.  In this paper, we present the first theoretical predictions of pull, focusing on a color-singlet decaying in two jets.  The pull angle observable is particularly sensitive to color flow, but is not infrared and collinear safe and so cannot be calculated in fixed-order perturbation theory.  Nevertheless, all-orders resummation renders its distribution finite, a property referred to as Sudakov safety.  In our prediction of the pull angle we also include an estimation of the effects from hadronization, and directly compare our results to simulation and experimental data.
\end{abstract}

\maketitle

Determining the short-distance origin of jets, manifestations of high energy quantum chromodynamics (QCD), is a problem of foremost importance at the Large Hadron Collider (LHC).  Some quantum numbers, such as the mass or electric charge, are relatively straightforward to measure at the LHC.  Determining the color representation of a jet or collection of jets, however, is highly non-trivial because the particles that carry color quantum numbers, quarks and gluons, are not directly observable in experiment.  The color representation must be inferred through its effect on kinematic distributions.  The observable pull~\cite{Gallicchio:2010sw}, and derivative quantities, is a widely used observable sensitive to the color representation.  Pull is a two-dimensional vector that points in the direction of dominant energy flow about a jet of interest that is particularly useful for determining if two jets form a color-singlet dipole, i.e.\ whether they originate from the decay of resonance that carries no color, such as an electroweak boson.  In a color-singlet dipole, emissions lie between the ends of the dipole; therefore the pull vector would point along the line that connects the momentum vectors of the jets.

In this Letter, we present the first analytic predictions from first-principles QCD for the pull vector.  We focus on the calculation of the pull vector for color-singlet dipoles, as this is the case that has been studied experimentally in detail.  The most useful feature of the pull vector for studying color dipoles is the pull angle, which is the azimuthal angle about one of the jets in a pair with respect to the line connecting the jets.  Both D$\emptyset$ and ATLAS experiments have measured the pull angle in the boosted, hadronic decays of $W$ bosons from top quark decay~\cite{Abazov:2011vh,Aad:2015lxa,Aaboud:2018ibj}.
It has been found that state-of-the-art general-purpose Monte Carlo simulations provide an unsatisfactory description of the data, thus indicating the need for dedicated first-principle calculations in QCD.
 However, unlike most theoretically-studied observables for jet physics, the pull angle lacks the property of infrared and collinear (IRC) safety, and so its distribution cannot be calculated in the fixed-order perturbation theory of QCD. Nevertheless, it is Sudakov safe~\cite{Larkoski:2013paa,Larkoski:2014wba,Larkoski:2015lea}, in that its distribution is rendered finite by including all-orders resummation.  With the theoretical prediction for the pull angle in hand, we then compare to Monte Carlo simulation and experimental data.

The original definition of the pull vector $\vec t$ from~\cite{Gallicchio:2010sw} was as a two-dimensional vector in the plane of rapidity $y$ and azimuthal angle $\phi$.  The expression for the pull vector is 
\begin{equation}\label{eq:pullorigdef}
\vec t_\text{original} = \sum_{i\in J}\frac{p_{\perp i} |\vec r_i|}{p_{\perp J}}\vec r_i\,.
\end{equation}
Here, $i$ is a particle in the jet $J$ of interest and $p_{\perp i}$ is its transverse momentum with respect to the collision beam axis.  The vector $\vec r_i$ is the relative rapidity and azimuthal angle of the particle from the jet axis:
\begin{equation}
\vec r_i = (y_i-y_J,\phi_i-\phi_J)\,.
\end{equation}
As a weighted sum of particle locations, the pull vector points from the jet axis in the direction of dominant energy flow.  In this form, the pull vector is expressed in coordinates natural at a hadron collider, and has been used for the measurements at D$\emptyset$ and ATLAS, and for searches at CMS~\cite{Chatrchyan:2012xdj,Chatrchyan:2013zna,Chatrchyan:2014tja,Khachatryan:2014dea,Khachatryan:2016mdm}.

For the calculations in this paper, we use a modified version of the pull vector, which is identical to \Eq{eq:pullorigdef} for central jets in the collinear limit.  The definition we use is
\begin{equation}
\vec t_\text{modified} = \sum_{i\in J}\frac{E_i \sin^2\theta_i}{E_J}(\cos\phi_i,\sin\phi_i)\,.
\end{equation}
Here, $E_i$ is the energy of particle $i$, $\theta_i$ is its angle from the jet axis, and $\phi_i$ is the azimuthal angle about the jet axis. The angle $\phi_i$ is measured with respect to a fiducial jet direction.  This form is much more amenable to analytic calculations, and because the jet radii that we consider are typically relatively small ($R \simeq 0.4$), the collinear limit is a good approximation anyway.  
To correct for the difference between the original definition which is used in experiment and this modified definition, we could match our resummed calculations to fixed-order results which would account for the difference. In what follows, we will refer to this version of the pull vector as $\vec t$, for brevity.

As a two-dimensional vector, $\vec t$ can be defined by a magnitude $|\vec t|\equiv t$ and an angle $\phi_p$.  When measured with respect to the line connecting the momentum vectors of two jets, $\phi_p$ is the pull angle observable.  The pull vector magnitude $t$ is itself IRC safe, and so can be calculated to any fixed-order.  While the pull angle $\phi_p$ is not IRC safe, the problematic region of phase space is localized to $t=0$, where the complete cross section vanishes anyway.  This motivates the calculation of the distribution of the pull angle $p(\phi_p)$ by marginalization of a joint probability distribution of $t$ and $\phi_p$:
\begin{equation}
p(\phi_p) = \int dt\, p(t,\phi_p)\,.
\end{equation} 
This only exists if the joint distribution is integrable, which is not true when calculated at fixed-order.  Following \Ref{Larkoski:2015lea}, we can make progress by re-expressing the joint distribution in terms of a conditional probability:
\begin{equation}
p(\phi_p) = \int dt\, p(t,\phi_p) = \int dt\, p(t)\,p(\phi_p|t)\,,
\end{equation} 
where $p(\phi_p|t)$ is the distribution of $\phi_p$ conditioned on the value of $t$.  $p(\phi_p|t)$ is finite for $t\neq 0$, so can be calculated to any fixed-order, while $p(t)$ is finite to any fixed-order and further can be calculated in resummed perturbation theory.  To render the integral finite, then, we resum $t$ and calculate the joint probability to fixed-order (fo):
 \begin{equation}
p(\phi_p) \simeq \int dt\, p_\text{resum}(t)\,p_\text{fo}(\phi_p|t)\,.
\end{equation} 
While this relationship is no longer an exact equality, it nevertheless exists, is formally accurate to a fixed-order with $t\ll 1$, and is systematically improvable.  In the language of \cite{Larkoski:2015lea}, the pull vector magnitude $t$ is the safe companion of the pull angle $\phi_p$.  

We now calculate these two distributions, one resummed and one at fixed-order.  Starting with the conditional distribution, we note that
\begin{equation}
p_\text{fo}(\phi_p|t) = \frac{p_\text{fo}(t,\phi_p)}{p_\text{fo}(t)}\,,
\end{equation}
where everything is calculated to the same order in $\alpha_s$.  Further, the fixed-order distribution of the pull magnitude is just a marginalization of the joint distribution
\begin{equation}
p_\text{fo}(t) = \int_0^{2\pi}d\phi_p\, p_\text{fo}(t,\phi_p)\,,
\end{equation}
so we just need to calculate the joint distribution.  We will calculate $p_\text{fo}(t,\phi_p)$ to leading-order in $\alpha_s$, in the soft and collinear limits; that is, to leading order for $t\ll 1$.  The soft and collinear limits can be separated from one another with dimensional regularization and therefore be calculated separately.  

For soft gluon emission off of a $q\bar q$ dipole, which originates from an electroweak boson decay, the distribution can be calculated from:
\begin{align}
&p_s(t,\phi_p)\\
& = g^2 C_F \mu^{2\epsilon}\int [d^dk]_+ \, \frac{2n_1\cdot n_2}{k^+(n_2\cdot k)}\,\Theta\left(\tan^2\frac{R}{2}-\frac{k^+}{k^-}\right)\nonumber\\
&
\hspace{4cm}\times \delta\left(
t - \frac{k^+ k^-}{E_J k^0}
\right)\delta(\phi_p-\phi)\nonumber\,.
\end{align}
Here, $g$ is the QCD coupling, $C_F=4/3$ is the fundamental Casimir, $\mu$ is the dimensional regularization scale, and $[d^dk]_+$ is on-shell, positive-energy, $d=4-2\epsilon$ dimensional phase space.  The light-like four-vectors $n_1$ and $n_2$ specify the directions of the two jets of the dipole, and we have chosen to demand that the emission lies within an angle $R$ of jet direction $n_1$.  $E_J$ is the energy of jet 1, $k^0$ is the energy of the emission, $\phi$ is the azimuthal angle of the emission about jet 1 with respect to jet 2 and we use the shorthand notation
\begin{align}
&k^+ = k^0(1-\cos\theta_{1k})\,, &k^- = k^0(1+\cos\theta_{1k})\,.
\end{align}
$\theta_{1k}$ is the angle between the emission of jet 1.  The renormalized joint distribution for a soft emission is then
\begin{align}
p_s(t,\phi_p)&= \frac{\alpha_s C_F}{2\pi^2} \frac{1}{t} \left[2\log\frac{\mu\tan\frac{R}{2}}{tE_J\sin\phi_p}\right.\\
&
\hspace{0.5cm}
+2\cot \phi_p\,\, \tan^{-1}\frac{\frac{\tan\frac{R}{2}}{\tan\frac{\theta_{12}}{2}}\sin\phi_p}{1-\frac{\tan\frac{R}{2}}{\tan\frac{\theta_{12}}{2}}\cos\phi_p}\nonumber\\
&
\left.
\hspace{0.5cm}-\log\left(
1+\frac{\tan^2\frac{R}{2}}{\tan^2\frac{\theta_{12}}{2}}-2\frac{\tan\frac{R}{2}}{\tan\frac{\theta_{12}}{2}}\cos\phi_p
\right) \right]\nonumber\,.
\end{align}
In this expression, $\theta_{12}$ is the angle between the two ends of the dipole, 
\begin{equation}
n_1\cdot n_2 = 1-\cos\theta_{12}\,.
\end{equation}

The collinear contribution to the joint distribution can be calculated similarly, with appropriate changes to the form of the pull vector observable in this limit.  For a collinear splitting of jet 1 in which one particle takes an energy fraction $z$ and the other $1-z$, the distribution can be calculated from
\begin{align}
&p_c(t,\phi_p)\\
& = g^2 \mu^{2\epsilon} \int [d^dk_1]_+[d^dk_2]_+\, (2\pi)^4\delta^{(4)}(p_J - k_1-k_2)\nonumber\\
&\hspace{1.5cm}
\times P_{qg\leftarrow q}(z)\,\delta(t-z(1-z)|1-2z|\theta^2)\delta(\phi_p-\phi)\nonumber\,.
\end{align}
Here, $p_J$ is the total four-vector of the jet which undergoes the splitting to particles with momenta $k_1$ and $k_2$, $P_{qg\leftarrow q}(z)$ is the collinear splitting function, and $\theta$ is the angle between the particles in the splitting
\begin{equation}
\frac{k_1\cdot k_2}{k^0_1k^0_2} = 1-\cos\theta\,.
\end{equation}
The peculiar form of the pull magnitude, with the factor $|1-2z|$, comes from the fact that the two particles in the splitting preserve the momentum of the jet, and so their azimuthal angles $\phi_1$ and $\phi_2$ about the jet axis must differ by $\pi$.  The renormalized collinear contribution to the joint distribution is then
\begin{align}
p_c(t,\phi_p) = \frac{\alpha_s C_F}{4\pi^2}\frac{1}{t}\left[
\log\frac{4tE_J^2\sin^2\phi_p}{\mu^2}-\frac{3}{2}
\right]\,.
\end{align}

Combining these soft and collinear results produces a lowest-order distribution that is independent of renormalization scale $\mu$:
\begin{align}
p_\text{fo}(t,\phi_p) &= p_s(t,\phi_p)+p_c(t,\phi_p)\\
&=\frac{\alpha_s C_F}{2\pi^2}\frac{1}{t}\left[\log\frac{4\tan^2\frac{R}{2}}{t}-\frac{3}{4}
\right.\nonumber\\
&
\hspace{1cm}
+2\cot \phi_p\,\, \tan^{-1}\frac{\frac{\tan\frac{R}{2}}{\tan\frac{\theta_{12}}{2}}\sin\phi_p}{1-\frac{\tan\frac{R}{2}}{\tan\frac{\theta_{12}}{2}}\cos\phi_p}\nonumber\\
&
\left.
\hspace{1cm}-\log\left(
1+\frac{\tan^2\frac{R}{2}}{\tan^2\frac{\theta_{12}}{2}}-2\frac{\tan\frac{R}{2}}{\tan\frac{\theta_{12}}{2}}\cos\phi_p
\right) \right]\nonumber\,.
\end{align}
By integrating over the pull angle $\phi_p$ we find the distribution of the pull magnitude $t$:
\begin{equation}
p_\text{fo}(t) = \frac{\alpha_s C_F}{\pi}\frac{1}{t}\left[
\log\frac{1}{t}-\frac{3}{4}-\log\left(\frac{1-\frac{\tan^2\frac{R}{2}}{\tan^2\frac{\theta_{12}}{2}}}{4\tan^2\frac{R}{2}}\right)
\right]\,.
\end{equation}
The ratio of these distributions then defines the fixed-order, conditional distribution, $p_\text{fo}(\phi_p|t)$.  Note that, to lowest order, this conditional probability distribution will be independent of the value of the coupling, $\alpha_s$.

Now, we calculate the resummed distribution of the pull magnitude $p_\text{resum}(t)$.  
Because experimental analyses typically consider a $W$ boson decaying into subjets with small radius, we decide to perform the all-order calculation in the collinear limit, although determine the full-$R$ dependence is a straightforward extension.  
We note that in this limit our definition of pull and the original one coincide.
The pull vector $\vec t$ is an additive observable in that the contribution to the pull vector from additional soft emissions simply add.  The pull vector is recoil-free in the sense that soft emissions do not affect the direction of the jet axis to leading power in the pull magnitude $t\ll1$.  With these observations, to next-to-leading logarithmic accuracy (NLL) in the collinear limit, the double differential cross section for the pull vector can be directly calculated from an infinite sum of jets with any number of emissions of energy fraction $\{z_i\}$ and emission angles~$\{\theta_i\}$:
\begin{align}\label{eq:t-resum}
&\frac{1}{\sigma}\frac{d^2\sigma}{d\vec t} = \exp\left[
-\int_0^{R^2}\frac{d\theta^2}{\theta^2}\int_0^1 dz \int_0^{2\pi}\frac{d\phi}{2\pi}\frac{\alpha_s}{2\pi}P_{qg\leftarrow q}(z)
\right]\nonumber \\
&
\times\left[
\sum_{n=0}^\infty\frac{1}{n!} \prod_{i=1}^n \int_0^{R^2}\frac{d\theta_i^2}{\theta_i^2}\int_0^1 dz_i \int_0^{2\pi}\frac{d\phi_i}{2\pi}\frac{\alpha_s}{2\pi}P_{qg\leftarrow q}(z_i)\right.\nonumber\\
&
\left.
\times\, \delta\left(t_x-\sum_{i=1}^n z_i \theta_i^2\cos\phi_i\right)\delta\left(t_y-\sum_{i=1}^n z_i \theta_i^2\sin\phi_i\right)
\phantom{int_0^{2\pi}}
\hspace{-0.8cm}\right]\,.
\end{align}
Momentum conservation of the collinear emissions has been suppressed, and this expression ignores non-global logarithms \cite{Dasgupta:2001sh} and powers of the jet radius $R$.

The structure of the resummed results is akin to the well-known transverse-momentum resummation, e.g.~\cite{Parisi:1979se,Collins:1984kg}, and consequently the sum can be done explicitly with a two-dimensional Fourier transform and the cross section can be expressed as
\begin{equation}\label{eq:two-dim-distr}
\frac{1}{\sigma}\frac{d^2\sigma}{d\vec t} = \int \frac{d^2 b}{(2\pi)^2}e^{i\vec b\cdot \vec t} e^{-R(b)}\,.
\end{equation}
$R(b)$ is the radiator, which, at this accuracy, depends exclusively on the magnitude of the Fourier conjugate vector $b=|\vec b|$:
\begin{align}
R(b) = \int_0^{R^2}\frac{d\theta^2}{\theta^2}\int_0^1 dz \frac{\alpha_s}{2\pi}P_{qg\leftarrow q}(z)\left(
1-J_0(bz\theta^2)
\right)\,,
\end{align}
where $J_0(x)$ is the Bessel function.  To determine the distribution for the magnitude of the pull vector $p_\text{resum}(t)$, we simply integrate over the pull angle $\phi_p$ and the $b$-space azimuthal angle to find:
\begin{align}
p_\text{resum}(t) =\frac{1}{\sigma}\frac{d \sigma}{d t}= t\int_0^\infty db\, bJ_0(bt)e^{-R(b)}\,.
\end{align}
This expression can be explicitly expanded and evaluated to NLL with the two-loop running coupling, but we leave it implicit here. 

There are two more things we include in our theoretical prediction of the pull angle.  First, in the calculation of the fixed-order conditional distribution $p_\text{fo}(t,\phi_p)$, there is explicit dependence on the angle between the two ends of the color singlet dipole, $\theta_{12}$.  Our expression for $p_\text{fo}(t,\phi_p)$, then, needs to be convolved against the distribution of this angle.  For isotropic color-singlet decays, this distribution can be determined by boosting the rest frame decay to the lab frame.  
We find
\begin{align}
&p(\cos \theta_{12}) = \frac{1}{\sqrt{\gamma^4-\gamma^2}}\frac{1}{\sqrt{1-\frac{2}{\gamma^2}-\cos\theta_{12}}}\frac{1}{(1-\cos\theta_{12})^{3/2}}\nonumber\\
&
\hspace{2cm} \times\Theta\left(
1-\frac{2}{\gamma^2}-\cos\theta_{12}
\right)\,,
\end{align}
where $\gamma$ is the boost factor.  For comparison with data, we need to integrate over all possible subjet angles accepted by the experimental cuts.  From ATLAS's analysis, most of the top quarks will be produced at or near rest, and so the $W$ boson's boost factor is approximately
\begin{equation}
\gamma = \frac{m_t^2+m_W^2}{2m_t m_W}\simeq 1.3\,.
\end{equation}
ATLAS also requires that the jets on which the pull angle is calculated have a minimum transverse momentum of $p_{\perp,\min}=25$ GeV, so assuming a purely transverse decay, the maximum angle between the jets is
\begin{equation}
\cos\theta_{12}\gtrsim 1-\frac{m_W^2}{2 p_{\perp,\min}(\gamma m_W-p_{\perp,\min})} \simeq -0.62\,.
\end{equation}
We use these parameters to form our complete theory prediction.  

The final component of our theory prediction is the inclusion of non-perturbative corrections from hadronization.  Due to the additivity of the pull vector, hadronization corrections can be included to leading power by convolution of the perturbative distribution with a model shape function \cite{Korchemsky:1999kt,Korchemsky:2000kp,Bosch:2004th,Hoang:2007vb,Ligeti:2008ac}.  This shape function encodes the kinematic distribution of non-perturbative emissions on which the pull vector is measured and is peaked around energies comparable to the QCD scale, $\Lambda_\text{QCD}$.  While one can do something more sophisticated, we parametrize the non-perturbative distribution of the pull vector as:
\begin{equation}
p_\text{np}(t,\phi_p) \propto \tanh\left(\frac{1}{a\phi_p(2\pi-\phi_p)}\right) \,\delta\left(t - \frac{\Omega}{E_J}\right)\,,
\end{equation}
where the constant of proportionality is defined by normalization.  Here, $\Omega \simeq \Lambda_\text{QCD}$, and the functional dependence of the pull angle $\phi_p$ has a free parameter $a$ for which $a\to 0$ yields a flat distribution in $\phi_p$ and $a\to \infty$ is a $\delta$-function at $\phi_p = 0$.  This form of the non-perturbative distribution is motivated by noting that in the center-of-mass frame of the color-singlet decay, at lowest order emissions are uniform in azimuth about the decay axis.  When boosted to the lab frame this naturally clusters emissions at small values of $\phi_p$.  We  find that varying the parameter $a \in \left[0,\frac{1}{4} \right]$ is sufficient in order to estimate the dependence on the precise shape of non-perturbative corrections.

With all of these pieces in place, we can then state the complete expression for the theoretical prediction of the pull angle distribution.  Step-by-step, the perturbative joint distribution $p_\text{perp}(t,\phi_p)$ of the pull magnitude and angle is
\begin{align}
&p_\text{perp}(t,\phi_p) \nonumber\\
&
\hspace{0.4cm}= \int_{-0.62}^{-0.18} d\cos\theta_{12}\, p_\text{resum}(t) \, p_\text{fo}(\phi_p|t) \, p(\cos\theta_{12})\,,
\end{align}
where the integration bounds follow from the earlier discussion of the boost of the $W$ boson in the lab frame.  Non-perturbative corrections can be included by convolution carefully vectorially summing the components of the pull vector:
\begin{align}
&p(t,\phi_p) \\
&
\hspace{-0.15cm}
= \int_0^\infty dt'\int_0^{2\pi}d\phi'\int_0^\infty dt''\int_0^{2\pi}d\phi''  p_\text{perp}(t',\phi') \, p_\text{np}(t'',\phi'')\nonumber\\
&
\hspace{2cm}\times\delta\left(
\phi_p - \cos^{-1}\frac{t'\cos\phi'+t''\cos\phi''}{t}
\right)\nonumber\\
&
\hspace{2cm}
\times\delta\left(t-\sqrt{t'^2+t''^2+2t' t''\cos(\phi'-\phi'')}\right)\,.\nonumber
\end{align}
Finally, integrating over the pull magnitude yields the pull angle distribution:
\begin{align}
p(\phi_p) = \int_0^\infty dt\, p(t,\phi_p)\,.
\end{align}

\begin{figure*}[t!]
\begin{center}
\includegraphics[width=0.48\textwidth]{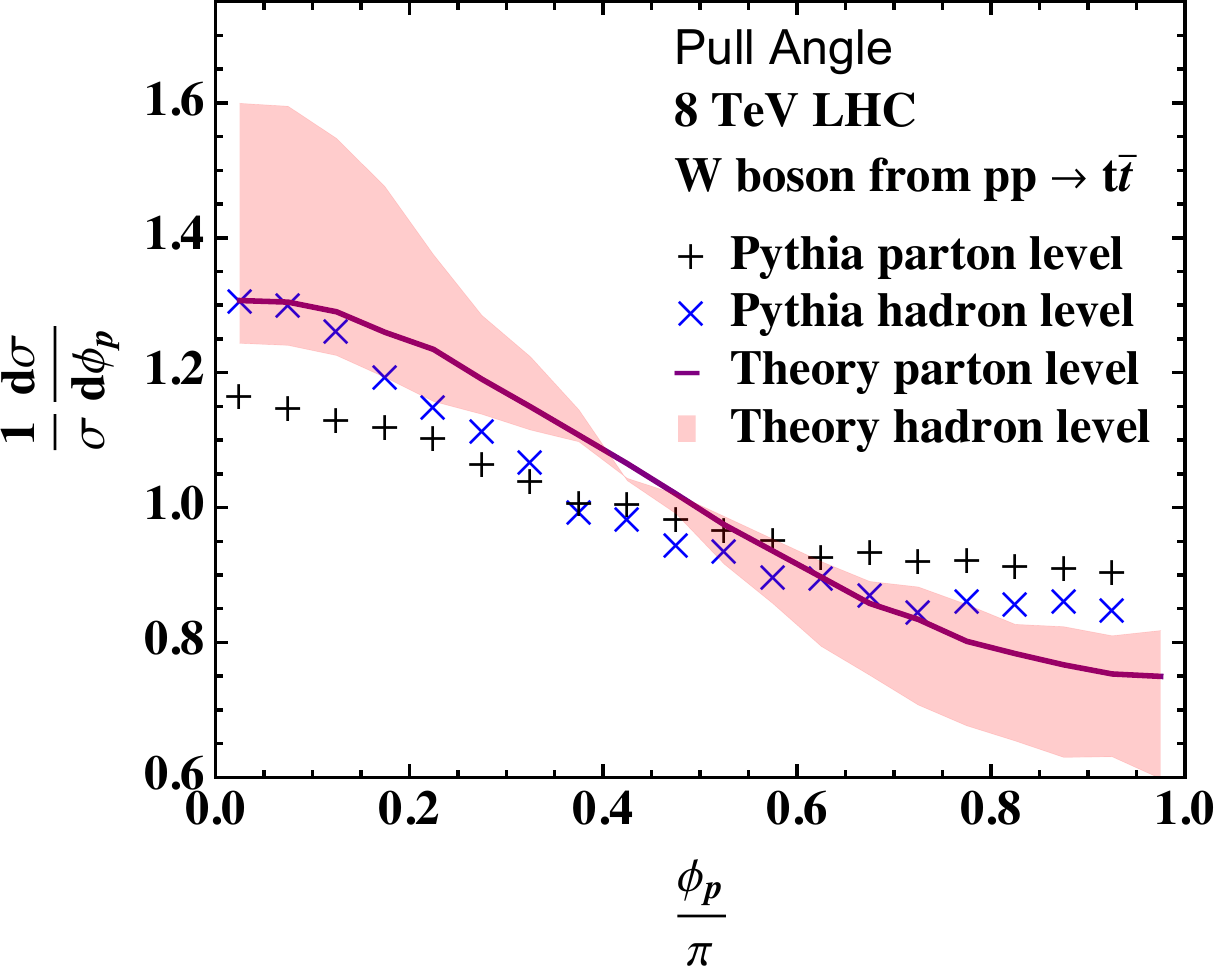}\hfill
\includegraphics[width=0.48\textwidth]{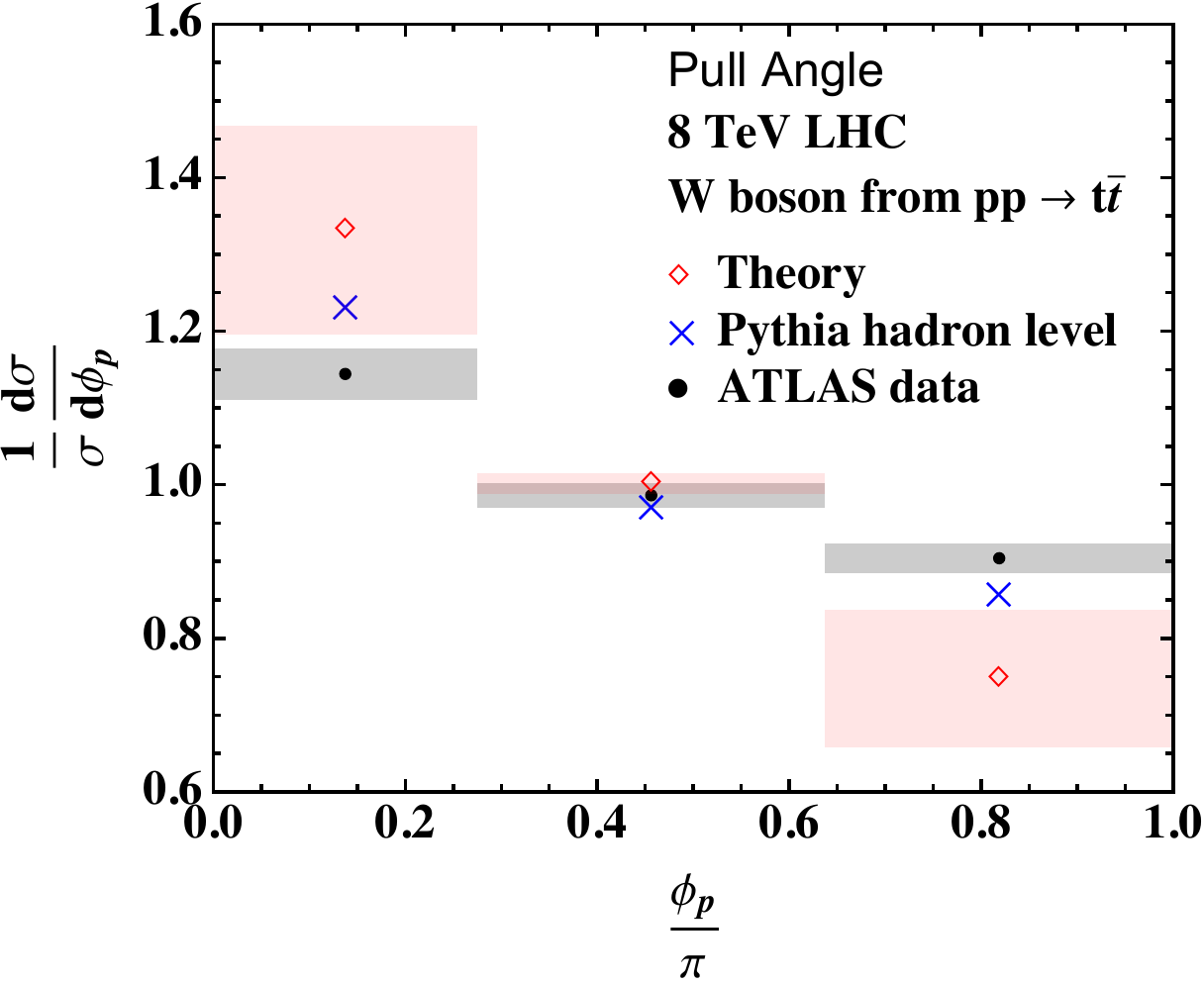}
\end{center}
\caption{In the left plot we show the distribution of the pull angle in various approximations. Monte Carlo simulations from Pythia are shown at parton level (black $+$) and hadron level (blue $\times$). Our theory prediction is also shown at parton level (purple line) and with varying hadronization corrections (light-red band). In the right plot we compare our hadron-level result (red $\diamond$) to the ATLAS data (black $\bullet$), as well as to the Monte Carlo simulation (blue $\times$).}
\label{fig:results}
\end{figure*}

Our theoretical prediction is plotted in \Fig{fig:results}. On the left-hand side, we show the pull distribution as computed in perturbative QCD and with hadronization corrections as described above.  
At small $\phi_p$, the lower edge of the band corresponds to $a=0$, while the upper one to $a=\frac{1}{4}$.
For comparison we also show simulated data both at parton level and hadron level. 
To produce the simulated events, we follow the experimental analysis of~\Ref{Aad:2015lxa}, where the pull angle measured on all particles from the two jets from hadronic $W$ decay in semi-leptonic $t\bar t$ events. Therefore we generate semi-leptonic $pp\to t\bar t$ events at the 8 TeV LHC with MadGraph v2.6.4~\cite{Alwall:2014hca} and then showered in Pythia v8.240 \cite{Sjostrand:2014zea}. FastJet v3.3.2 \cite{Cacciari:2011ma} was used to impose phase space restrictions from the ATLAS analysis, find jets, and calculate the pull angle.  
Finally, on the right-hand side \Fig{fig:results} we compare theory and Monte Carlo predictions at hadron-level to data collected by the ATLAS experiment of \Ref{Aad:2015lxa}, which are available from HEPData \cite{1376945}. As central value of our hadron-level theoretical prediction we consider the midpoint of the $0<a<\frac{1}{4}$ band.
We note that both theory calculation and simulation predict a distribution of the pull angle that is slightly more peaked at small values than data, which was also observed in ATLAS's analysis.

It is known that the pull angle shows more sensitivity to color flow than the pull magnitude itself, but it has the drawback of being IRC unsafe. Despite the fact that we were able to obtain a first-principle description of its distribution exploiting its Sudakov safety, it would be better to employ an IRC safe observable that maintains the same sensitivity. 
The projection of the pull vector along line connecting the two jets, i.e.\ the variable $t_x$ in \Eq{eq:t-resum}, is IRC safe and enjoys many of the same properties as the pull angle. Exploiting once again the similarities with transverse-momentum resummation, this observable can be resummed using the techniques developed in~\cite{Banfi:2009dy,Banfi:2011dx}. We are looking forward to future work on this topic. 

\newpage
\begin{acknowledgments}
We thank Ben Nachman, Yvonne Peters, Matthew Schwartz, Gregory Soyez, and Jesse Thaler for useful discussions and comments. 
This work is partly supported by the curiosity-driven grant ``Using jets to challenge the Standard Model of particle physics" from Universit\`a di Genova. 
\end{acknowledgments}

\bibliography{pull}

\begin{thebibliography}{26}%
\makeatletter
\providecommand \@ifxundefined [1]{%
 \@ifx{#1\undefined}
}%
\providecommand \@ifnum [1]{%
 \ifnum #1\expandafter \@firstoftwo
 \else \expandafter \@secondoftwo
 \fi
}%
\providecommand \@ifx [1]{%
 \ifx #1\expandafter \@firstoftwo
 \else \expandafter \@secondoftwo
 \fi
}%
\providecommand \natexlab [1]{#1}%
\providecommand \enquote  [1]{``#1''}%
\providecommand \bibnamefont  [1]{#1}%
\providecommand \bibfnamefont [1]{#1}%
\providecommand \citenamefont [1]{#1}%
\providecommand \href@noop [0]{\@secondoftwo}%
\providecommand \href [0]{\begingroup \@sanitize@url \@href}%
\providecommand \@href[1]{\@@startlink{#1}\@@href}%
\providecommand \@@href[1]{\endgroup#1\@@endlink}%
\providecommand \@sanitize@url [0]{\catcode `\\12\catcode `\$12\catcode
  `\&12\catcode `\#12\catcode `\^12\catcode `\_12\catcode `\%12\relax}%
\providecommand \@@startlink[1]{}%
\providecommand \@@endlink[0]{}%
\providecommand \url  [0]{\begingroup\@sanitize@url \@url }%
\providecommand \@url [1]{\endgroup\@href {#1}{\urlprefix }}%
\providecommand \urlprefix  [0]{URL }%
\providecommand \Eprint [0]{\href }%
\providecommand \doibase [0]{http://dx.doi.org/}%
\providecommand \selectlanguage [0]{\@gobble}%
\providecommand \bibinfo  [0]{\@secondoftwo}%
\providecommand \bibfield  [0]{\@secondoftwo}%
\providecommand \translation [1]{[#1]}%
\providecommand \BibitemOpen [0]{}%
\providecommand \bibitemStop [0]{}%
\providecommand \bibitemNoStop [0]{.\EOS\space}%
\providecommand \EOS [0]{\spacefactor3000\relax}%
\providecommand \BibitemShut  [1]{\csname bibitem#1\endcsname}%
\let\auto@bib@innerbib\@empty
\bibitem [{\citenamefont {Gallicchio}\ and\ \citenamefont
  {Schwartz}(2010)}]{Gallicchio:2010sw}%
  \BibitemOpen
  \bibfield  {author} {\bibinfo {author} {\bibfnamefont {Jason}\ \bibnamefont
  {Gallicchio}}\ and\ \bibinfo {author} {\bibfnamefont {Matthew~D.}\
  \bibnamefont {Schwartz}},\ }\bibfield  {title} {\enquote {\bibinfo {title}
  {{Seeing in Color: Jet Superstructure}},}\ }\href {\doibase
  10.1103/PhysRevLett.105.022001} {\bibfield  {journal} {\bibinfo  {journal}
  {Phys. Rev. Lett.}\ }\textbf {\bibinfo {volume} {105}},\ \bibinfo {pages}
  {022001} (\bibinfo {year} {2010})},\ \Eprint {http://arxiv.org/abs/1001.5027}
  {arXiv:1001.5027 [hep-ph]} \BibitemShut {NoStop}%
\bibitem [{\citenamefont {Abazov}\ \emph {et~al.}(2011)\citenamefont {Abazov}
  \emph {et~al.}}]{Abazov:2011vh}%
  \BibitemOpen
  \bibfield  {author} {\bibinfo {author} {\bibfnamefont {Victor~Mukhamedovich}\
  \bibnamefont {Abazov}} \emph {et~al.} (\bibinfo {collaboration} {D0}),\
  }\bibfield  {title} {\enquote {\bibinfo {title} {{Measurement of color flow
  in $\mathbf{t\bar{t}}$ events from $\mathbf{p\bar{p}}$ collisions at
  $\mathbf{\sqrt{s}=1.96}$ TeV}},}\ }\href {\doibase
  10.1103/PhysRevD.83.092002} {\bibfield  {journal} {\bibinfo  {journal} {Phys.
  Rev.}\ }\textbf {\bibinfo {volume} {D83}},\ \bibinfo {pages} {092002}
  (\bibinfo {year} {2011})},\ \Eprint {http://arxiv.org/abs/1101.0648}
  {arXiv:1101.0648 [hep-ex]} \BibitemShut {NoStop}%
\bibitem [{\citenamefont {Aad}\ \emph {et~al.}(2015)\citenamefont {Aad} \emph
  {et~al.}}]{Aad:2015lxa}%
  \BibitemOpen
  \bibfield  {author} {\bibinfo {author} {\bibfnamefont {Georges}\ \bibnamefont
  {Aad}} \emph {et~al.} (\bibinfo {collaboration} {ATLAS}),\ }\bibfield
  {title} {\enquote {\bibinfo {title} {{Measurement of colour flow with the jet
  pull angle in $t\bar{t}$ events using the ATLAS detector at $\sqrt{s}=8$
  TeV}},}\ }\href {\doibase 10.1016/j.physletb.2015.09.051} {\bibfield
  {journal} {\bibinfo  {journal} {Phys. Lett.}\ }\textbf {\bibinfo {volume}
  {B750}},\ \bibinfo {pages} {475--493} (\bibinfo {year} {2015})},\ \Eprint
  {http://arxiv.org/abs/1506.05629} {arXiv:1506.05629 [hep-ex]} \BibitemShut
  {NoStop}%
\bibitem [{\citenamefont {Aaboud}\ \emph {et~al.}(2018)\citenamefont {Aaboud}
  \emph {et~al.}}]{Aaboud:2018ibj}%
  \BibitemOpen
  \bibfield  {author} {\bibinfo {author} {\bibfnamefont {Morad}\ \bibnamefont
  {Aaboud}} \emph {et~al.} (\bibinfo {collaboration} {ATLAS}),\ }\bibfield
  {title} {\enquote {\bibinfo {title} {{Measurement of colour flow using
  jet-pull observables in $t\bar{t}$ events with the ATLAS experiment at
  $\sqrt{s} = 13\,\hbox {TeV}$}},}\ }\href {\doibase
  10.1140/epjc/s10052-018-6290-2} {\bibfield  {journal} {\bibinfo  {journal}
  {Eur. Phys. J.}\ }\textbf {\bibinfo {volume} {C78}},\ \bibinfo {pages} {847}
  (\bibinfo {year} {2018})},\ \Eprint {http://arxiv.org/abs/1805.02935}
  {arXiv:1805.02935 [hep-ex]} \BibitemShut {NoStop}%
\bibitem [{\citenamefont {Larkoski}\ and\ \citenamefont
  {Thaler}(2013)}]{Larkoski:2013paa}%
  \BibitemOpen
  \bibfield  {author} {\bibinfo {author} {\bibfnamefont {Andrew~J.}\
  \bibnamefont {Larkoski}}\ and\ \bibinfo {author} {\bibfnamefont {Jesse}\
  \bibnamefont {Thaler}},\ }\bibfield  {title} {\enquote {\bibinfo {title}
  {{Unsafe but Calculable: Ratios of Angularities in Perturbative QCD}},}\
  }\href {\doibase 10.1007/JHEP09(2013)137} {\bibfield  {journal} {\bibinfo
  {journal} {JHEP}\ }\textbf {\bibinfo {volume} {1309}},\ \bibinfo {pages}
  {137} (\bibinfo {year} {2013})},\ \Eprint {http://arxiv.org/abs/1307.1699}
  {arXiv:1307.1699} \BibitemShut {NoStop}%
\bibitem [{\citenamefont {Larkoski}\ \emph {et~al.}(2014)\citenamefont
  {Larkoski}, \citenamefont {Marzani}, \citenamefont {Soyez},\ and\
  \citenamefont {Thaler}}]{Larkoski:2014wba}%
  \BibitemOpen
  \bibfield  {author} {\bibinfo {author} {\bibfnamefont {Andrew~J.}\
  \bibnamefont {Larkoski}}, \bibinfo {author} {\bibfnamefont {Simone}\
  \bibnamefont {Marzani}}, \bibinfo {author} {\bibfnamefont {Gregory}\
  \bibnamefont {Soyez}}, \ and\ \bibinfo {author} {\bibfnamefont {Jesse}\
  \bibnamefont {Thaler}},\ }\bibfield  {title} {\enquote {\bibinfo {title}
  {{Soft Drop}},}\ }\href {\doibase 10.1007/JHEP05(2014)146} {\bibfield
  {journal} {\bibinfo  {journal} {JHEP}\ }\textbf {\bibinfo {volume} {1405}},\
  \bibinfo {pages} {146} (\bibinfo {year} {2014})},\ \Eprint
  {http://arxiv.org/abs/1402.2657} {arXiv:1402.2657 [hep-ph]} \BibitemShut
  {NoStop}%
\bibitem [{\citenamefont {Larkoski}\ \emph {et~al.}(2015)\citenamefont
  {Larkoski}, \citenamefont {Marzani},\ and\ \citenamefont
  {Thaler}}]{Larkoski:2015lea}%
  \BibitemOpen
  \bibfield  {author} {\bibinfo {author} {\bibfnamefont {Andrew~J.}\
  \bibnamefont {Larkoski}}, \bibinfo {author} {\bibfnamefont {Simone}\
  \bibnamefont {Marzani}}, \ and\ \bibinfo {author} {\bibfnamefont {Jesse}\
  \bibnamefont {Thaler}},\ }\bibfield  {title} {\enquote {\bibinfo {title}
  {{Sudakov Safety in Perturbative QCD}},}\ }\href {\doibase
  10.1103/PhysRevD.91.111501} {\bibfield  {journal} {\bibinfo  {journal}
  {Phys.Rev.}\ }\textbf {\bibinfo {volume} {D91}},\ \bibinfo {pages} {111501}
  (\bibinfo {year} {2015})},\ \Eprint {http://arxiv.org/abs/1502.01719}
  {arXiv:1502.01719 [hep-ph]} \BibitemShut {NoStop}%
\bibitem [{\citenamefont {Chatrchyan}\ \emph {et~al.}(2012)\citenamefont
  {Chatrchyan} \emph {et~al.}}]{Chatrchyan:2012xdj}%
  \BibitemOpen
  \bibfield  {author} {\bibinfo {author} {\bibfnamefont {Serguei}\ \bibnamefont
  {Chatrchyan}} \emph {et~al.} (\bibinfo {collaboration} {CMS}),\ }\bibfield
  {title} {\enquote {\bibinfo {title} {{Observation of a new boson at a mass of
  125 GeV with the CMS experiment at the LHC}},}\ }\href {\doibase
  10.1016/j.physletb.2012.08.021} {\bibfield  {journal} {\bibinfo  {journal}
  {Phys. Lett.}\ }\textbf {\bibinfo {volume} {B716}},\ \bibinfo {pages}
  {30--61} (\bibinfo {year} {2012})},\ \Eprint {http://arxiv.org/abs/1207.7235}
  {arXiv:1207.7235 [hep-ex]} \BibitemShut {NoStop}%
\bibitem [{\citenamefont {Chatrchyan}\ \emph
  {et~al.}(2014{\natexlab{a}})\citenamefont {Chatrchyan} \emph
  {et~al.}}]{Chatrchyan:2013zna}%
  \BibitemOpen
  \bibfield  {author} {\bibinfo {author} {\bibfnamefont {Serguei}\ \bibnamefont
  {Chatrchyan}} \emph {et~al.} (\bibinfo {collaboration} {CMS}),\ }\bibfield
  {title} {\enquote {\bibinfo {title} {{Search for the standard model Higgs
  boson produced in association with a W or a Z boson and decaying to bottom
  quarks}},}\ }\href {\doibase 10.1103/PhysRevD.89.012003} {\bibfield
  {journal} {\bibinfo  {journal} {Phys. Rev.}\ }\textbf {\bibinfo {volume}
  {D89}},\ \bibinfo {pages} {012003} (\bibinfo {year} {2014}{\natexlab{a}})},\
  \Eprint {http://arxiv.org/abs/1310.3687} {arXiv:1310.3687 [hep-ex]}
  \BibitemShut {NoStop}%
\bibitem [{\citenamefont {Chatrchyan}\ \emph
  {et~al.}(2014{\natexlab{b}})\citenamefont {Chatrchyan} \emph
  {et~al.}}]{Chatrchyan:2014tja}%
  \BibitemOpen
  \bibfield  {author} {\bibinfo {author} {\bibfnamefont {Serguei}\ \bibnamefont
  {Chatrchyan}} \emph {et~al.} (\bibinfo {collaboration} {CMS}),\ }\bibfield
  {title} {\enquote {\bibinfo {title} {{Search for invisible decays of Higgs
  bosons in the vector boson fusion and associated ZH production modes}},}\
  }\href {\doibase 10.1140/epjc/s10052-014-2980-6} {\bibfield  {journal}
  {\bibinfo  {journal} {Eur. Phys. J.}\ }\textbf {\bibinfo {volume} {C74}},\
  \bibinfo {pages} {2980} (\bibinfo {year} {2014}{\natexlab{b}})},\ \Eprint
  {http://arxiv.org/abs/1404.1344} {arXiv:1404.1344 [hep-ex]} \BibitemShut
  {NoStop}%
\bibitem [{\citenamefont {Khachatryan}\ \emph {et~al.}(2015)\citenamefont
  {Khachatryan} \emph {et~al.}}]{Khachatryan:2014dea}%
  \BibitemOpen
  \bibfield  {author} {\bibinfo {author} {\bibfnamefont {Vardan}\ \bibnamefont
  {Khachatryan}} \emph {et~al.} (\bibinfo {collaboration} {CMS}),\ }\bibfield
  {title} {\enquote {\bibinfo {title} {{Measurement of electroweak production
  of two jets in association with a Z boson in proton-proton collisions at
  $\sqrt{s}=8\,\text {TeV}$}},}\ }\href {\doibase
  10.1140/epjc/s10052-014-3232-5} {\bibfield  {journal} {\bibinfo  {journal}
  {Eur. Phys. J.}\ }\textbf {\bibinfo {volume} {C75}},\ \bibinfo {pages} {66}
  (\bibinfo {year} {2015})},\ \Eprint {http://arxiv.org/abs/1410.3153}
  {arXiv:1410.3153 [hep-ex]} \BibitemShut {NoStop}%
\bibitem [{\citenamefont {Khachatryan}\ \emph {et~al.}(2016)\citenamefont
  {Khachatryan} \emph {et~al.}}]{Khachatryan:2016mdm}%
  \BibitemOpen
  \bibfield  {author} {\bibinfo {author} {\bibfnamefont {Vardan}\ \bibnamefont
  {Khachatryan}} \emph {et~al.} (\bibinfo {collaboration} {CMS}),\ }\bibfield
  {title} {\enquote {\bibinfo {title} {{Search for dark matter in proton-proton
  collisions at 8 TeV with missing transverse momentum and vector boson tagged
  jets}},}\ }\href {\doibase 10.1007/JHEP12(2016)083, 10.1007/JHEP08(2017)035}
  {\bibfield  {journal} {\bibinfo  {journal} {JHEP}\ }\textbf {\bibinfo
  {volume} {12}},\ \bibinfo {pages} {083} (\bibinfo {year} {2016})},\ \bibinfo
  {note} {[Erratum: JHEP08,035(2017)]},\ \Eprint
  {http://arxiv.org/abs/1607.05764} {arXiv:1607.05764 [hep-ex]} \BibitemShut
  {NoStop}%
\bibitem [{\citenamefont {Dasgupta}\ and\ \citenamefont
  {Salam}(2001)}]{Dasgupta:2001sh}%
  \BibitemOpen
  \bibfield  {author} {\bibinfo {author} {\bibfnamefont {M.}~\bibnamefont
  {Dasgupta}}\ and\ \bibinfo {author} {\bibfnamefont {G.P.}\ \bibnamefont
  {Salam}},\ }\bibfield  {title} {\enquote {\bibinfo {title} {{Resummation of
  nonglobal QCD observables}},}\ }\href {\doibase
  10.1016/S0370-2693(01)00725-0} {\bibfield  {journal} {\bibinfo  {journal}
  {Phys.Lett.}\ }\textbf {\bibinfo {volume} {B512}},\ \bibinfo {pages}
  {323--330} (\bibinfo {year} {2001})},\ \Eprint
  {http://arxiv.org/abs/hep-ph/0104277} {arXiv:hep-ph/0104277 [hep-ph]}
  \BibitemShut {NoStop}%
\bibitem [{\citenamefont {Parisi}\ and\ \citenamefont
  {Petronzio}(1979)}]{Parisi:1979se}%
  \BibitemOpen
  \bibfield  {author} {\bibinfo {author} {\bibfnamefont {G.}~\bibnamefont
  {Parisi}}\ and\ \bibinfo {author} {\bibfnamefont {R.}~\bibnamefont
  {Petronzio}},\ }\bibfield  {title} {\enquote {\bibinfo {title} {{Small
  Transverse Momentum Distributions in Hard Processes}},}\ }\href {\doibase
  10.1016/0550-3213(79)90040-3} {\bibfield  {journal} {\bibinfo  {journal}
  {Nucl. Phys.}\ }\textbf {\bibinfo {volume} {B154}},\ \bibinfo {pages}
  {427--440} (\bibinfo {year} {1979})}\BibitemShut {NoStop}%
\bibitem [{\citenamefont {Collins}\ \emph {et~al.}(1985)\citenamefont
  {Collins}, \citenamefont {Soper},\ and\ \citenamefont
  {Sterman}}]{Collins:1984kg}%
  \BibitemOpen
  \bibfield  {author} {\bibinfo {author} {\bibfnamefont {John~C.}\ \bibnamefont
  {Collins}}, \bibinfo {author} {\bibfnamefont {Davison~E.}\ \bibnamefont
  {Soper}}, \ and\ \bibinfo {author} {\bibfnamefont {George~F.}\ \bibnamefont
  {Sterman}},\ }\bibfield  {title} {\enquote {\bibinfo {title} {{Transverse
  Momentum Distribution in Drell-Yan Pair and W and Z Boson Production}},}\
  }\href {\doibase 10.1016/0550-3213(85)90479-1} {\bibfield  {journal}
  {\bibinfo  {journal} {Nucl. Phys.}\ }\textbf {\bibinfo {volume} {B250}},\
  \bibinfo {pages} {199--224} (\bibinfo {year} {1985})}\BibitemShut {NoStop}%
\bibitem [{\citenamefont {Korchemsky}\ and\ \citenamefont
  {Sterman}(1999)}]{Korchemsky:1999kt}%
  \BibitemOpen
  \bibfield  {author} {\bibinfo {author} {\bibfnamefont {Gregory~P.}\
  \bibnamefont {Korchemsky}}\ and\ \bibinfo {author} {\bibfnamefont
  {George~F.}\ \bibnamefont {Sterman}},\ }\bibfield  {title} {\enquote
  {\bibinfo {title} {{Power corrections to event shapes and factorization}},}\
  }\href {\doibase 10.1016/S0550-3213(99)00308-9} {\bibfield  {journal}
  {\bibinfo  {journal} {Nucl.Phys.}\ }\textbf {\bibinfo {volume} {B555}},\
  \bibinfo {pages} {335--351} (\bibinfo {year} {1999})},\ \Eprint
  {http://arxiv.org/abs/hep-ph/9902341} {arXiv:hep-ph/9902341 [hep-ph]}
  \BibitemShut {NoStop}%
\bibitem [{\citenamefont {Korchemsky}\ and\ \citenamefont
  {Tafat}(2000)}]{Korchemsky:2000kp}%
  \BibitemOpen
  \bibfield  {author} {\bibinfo {author} {\bibfnamefont {G.P.}\ \bibnamefont
  {Korchemsky}}\ and\ \bibinfo {author} {\bibfnamefont {S.}~\bibnamefont
  {Tafat}},\ }\bibfield  {title} {\enquote {\bibinfo {title} {{On power
  corrections to the event shape distributions in QCD}},}\ }\href {\doibase
  10.1088/1126-6708/2000/10/010} {\bibfield  {journal} {\bibinfo  {journal}
  {JHEP}\ }\textbf {\bibinfo {volume} {0010}},\ \bibinfo {pages} {010}
  (\bibinfo {year} {2000})},\ \Eprint {http://arxiv.org/abs/hep-ph/0007005}
  {arXiv:hep-ph/0007005 [hep-ph]} \BibitemShut {NoStop}%
\bibitem [{\citenamefont {Bosch}\ \emph {et~al.}(2004)\citenamefont {Bosch},
  \citenamefont {Lange}, \citenamefont {Neubert},\ and\ \citenamefont
  {Paz}}]{Bosch:2004th}%
  \BibitemOpen
  \bibfield  {author} {\bibinfo {author} {\bibfnamefont {S.W.}\ \bibnamefont
  {Bosch}}, \bibinfo {author} {\bibfnamefont {B.O.}\ \bibnamefont {Lange}},
  \bibinfo {author} {\bibfnamefont {M.}~\bibnamefont {Neubert}}, \ and\
  \bibinfo {author} {\bibfnamefont {Gil}\ \bibnamefont {Paz}},\ }\bibfield
  {title} {\enquote {\bibinfo {title} {{Factorization and shape function
  effects in inclusive B meson decays}},}\ }\href {\doibase
  10.1016/j.nuclphysb.2004.07.041} {\bibfield  {journal} {\bibinfo  {journal}
  {Nucl.Phys.}\ }\textbf {\bibinfo {volume} {B699}},\ \bibinfo {pages}
  {335--386} (\bibinfo {year} {2004})},\ \Eprint
  {http://arxiv.org/abs/hep-ph/0402094} {arXiv:hep-ph/0402094 [hep-ph]}
  \BibitemShut {NoStop}%
\bibitem [{\citenamefont {Hoang}\ and\ \citenamefont
  {Stewart}(2008)}]{Hoang:2007vb}%
  \BibitemOpen
  \bibfield  {author} {\bibinfo {author} {\bibfnamefont {Andre~H.}\
  \bibnamefont {Hoang}}\ and\ \bibinfo {author} {\bibfnamefont {Iain~W.}\
  \bibnamefont {Stewart}},\ }\bibfield  {title} {\enquote {\bibinfo {title}
  {{Designing gapped soft functions for jet production}},}\ }\href {\doibase
  10.1016/j.physletb.2008.01.040} {\bibfield  {journal} {\bibinfo  {journal}
  {Phys.Lett.}\ }\textbf {\bibinfo {volume} {B660}},\ \bibinfo {pages}
  {483--493} (\bibinfo {year} {2008})},\ \Eprint
  {http://arxiv.org/abs/0709.3519} {arXiv:0709.3519 [hep-ph]} \BibitemShut
  {NoStop}%
\bibitem [{\citenamefont {Ligeti}\ \emph {et~al.}(2008)\citenamefont {Ligeti},
  \citenamefont {Stewart},\ and\ \citenamefont {Tackmann}}]{Ligeti:2008ac}%
  \BibitemOpen
  \bibfield  {author} {\bibinfo {author} {\bibfnamefont {Zoltan}\ \bibnamefont
  {Ligeti}}, \bibinfo {author} {\bibfnamefont {Iain~W.}\ \bibnamefont
  {Stewart}}, \ and\ \bibinfo {author} {\bibfnamefont {Frank~J.}\ \bibnamefont
  {Tackmann}},\ }\bibfield  {title} {\enquote {\bibinfo {title} {{Treating the
  b quark distribution function with reliable uncertainties}},}\ }\href
  {\doibase 10.1103/PhysRevD.78.114014} {\bibfield  {journal} {\bibinfo
  {journal} {Phys.Rev.}\ }\textbf {\bibinfo {volume} {D78}},\ \bibinfo {pages}
  {114014} (\bibinfo {year} {2008})},\ \Eprint {http://arxiv.org/abs/0807.1926}
  {arXiv:0807.1926 [hep-ph]} \BibitemShut {NoStop}%
\bibitem [{\citenamefont {Alwall}\ \emph {et~al.}(2014)\citenamefont {Alwall},
  \citenamefont {Frederix}, \citenamefont {Frixione}, \citenamefont {Hirschi},
  \citenamefont {Maltoni} \emph {et~al.}}]{Alwall:2014hca}%
  \BibitemOpen
  \bibfield  {author} {\bibinfo {author} {\bibfnamefont {J.}~\bibnamefont
  {Alwall}}, \bibinfo {author} {\bibfnamefont {R.}~\bibnamefont {Frederix}},
  \bibinfo {author} {\bibfnamefont {S.}~\bibnamefont {Frixione}}, \bibinfo
  {author} {\bibfnamefont {V.}~\bibnamefont {Hirschi}}, \bibinfo {author}
  {\bibfnamefont {F.}~\bibnamefont {Maltoni}},  \emph {et~al.},\ }\bibfield
  {title} {\enquote {\bibinfo {title} {{The automated computation of tree-level
  and next-to-leading order differential cross sections, and their matching to
  parton shower simulations}},}\ }\href {\doibase 10.1007/JHEP07(2014)079}
  {\bibfield  {journal} {\bibinfo  {journal} {JHEP}\ }\textbf {\bibinfo
  {volume} {1407}},\ \bibinfo {pages} {079} (\bibinfo {year} {2014})},\ \Eprint
  {http://arxiv.org/abs/1405.0301} {arXiv:1405.0301 [hep-ph]} \BibitemShut
  {NoStop}%
\bibitem [{\citenamefont {Sjostrand}\ \emph {et~al.}(2015)\citenamefont
  {Sjostrand}, \citenamefont {Ask}, \citenamefont {Christiansen}, \citenamefont
  {Corke}, \citenamefont {Desai}, \citenamefont {Ilten}, \citenamefont
  {Mrenna}, \citenamefont {Prestel}, \citenamefont {Rasmussen},\ and\
  \citenamefont {Skands}}]{Sjostrand:2014zea}%
  \BibitemOpen
  \bibfield  {author} {\bibinfo {author} {\bibfnamefont {Torbjorn}\
  \bibnamefont {Sjostrand}}, \bibinfo {author} {\bibfnamefont {Stefan}\
  \bibnamefont {Ask}}, \bibinfo {author} {\bibfnamefont {Jesper~R.}\
  \bibnamefont {Christiansen}}, \bibinfo {author} {\bibfnamefont {Richard}\
  \bibnamefont {Corke}}, \bibinfo {author} {\bibfnamefont {Nishita}\
  \bibnamefont {Desai}}, \bibinfo {author} {\bibfnamefont {Philip}\
  \bibnamefont {Ilten}}, \bibinfo {author} {\bibfnamefont {Stephen}\
  \bibnamefont {Mrenna}}, \bibinfo {author} {\bibfnamefont {Stefan}\
  \bibnamefont {Prestel}}, \bibinfo {author} {\bibfnamefont {Christine~O.}\
  \bibnamefont {Rasmussen}}, \ and\ \bibinfo {author} {\bibfnamefont
  {Peter~Z.}\ \bibnamefont {Skands}},\ }\bibfield  {title} {\enquote {\bibinfo
  {title} {{An Introduction to PYTHIA 8.2}},}\ }\href {\doibase
  10.1016/j.cpc.2015.01.024} {\bibfield  {journal} {\bibinfo  {journal}
  {Comput. Phys. Commun.}\ }\textbf {\bibinfo {volume} {191}},\ \bibinfo
  {pages} {159--177} (\bibinfo {year} {2015})},\ \Eprint
  {http://arxiv.org/abs/1410.3012} {arXiv:1410.3012 [hep-ph]} \BibitemShut
  {NoStop}%
\bibitem [{\citenamefont {Cacciari}\ \emph {et~al.}(2012)\citenamefont
  {Cacciari}, \citenamefont {Salam},\ and\ \citenamefont
  {Soyez}}]{Cacciari:2011ma}%
  \BibitemOpen
  \bibfield  {author} {\bibinfo {author} {\bibfnamefont {Matteo}\ \bibnamefont
  {Cacciari}}, \bibinfo {author} {\bibfnamefont {Gavin~P.}\ \bibnamefont
  {Salam}}, \ and\ \bibinfo {author} {\bibfnamefont {Gregory}\ \bibnamefont
  {Soyez}},\ }\bibfield  {title} {\enquote {\bibinfo {title} {{FastJet User
  Manual}},}\ }\href {\doibase 10.1140/epjc/s10052-012-1896-2} {\bibfield
  {journal} {\bibinfo  {journal} {Eur.Phys.J.}\ }\textbf {\bibinfo {volume}
  {C72}},\ \bibinfo {pages} {1896} (\bibinfo {year} {2012})},\ \Eprint
  {http://arxiv.org/abs/1111.6097} {arXiv:1111.6097 [hep-ph]} \BibitemShut
  {NoStop}%
\bibitem [{137(2015)}]{1376945}%
  \BibitemOpen
  \href@noop {} {\enquote {\bibinfo {title} {{Measurement of colour flow with
  the jet pull angle in $t\bar{t}$ events using the ATLAS detector at
  $\sqrt{s}=8$ TeV}},}\ }\bibinfo {howpublished}
  {\url{https://doi.org/10.17182/hepdata.71235}} (\bibinfo {year}
  {2015})\BibitemShut {NoStop}%
\bibitem [{\citenamefont {Banfi}\ \emph {et~al.}(2009)\citenamefont {Banfi},
  \citenamefont {Dasgupta},\ and\ \citenamefont
  {Duran~Delgado}}]{Banfi:2009dy}%
  \BibitemOpen
  \bibfield  {author} {\bibinfo {author} {\bibfnamefont {Andrea}\ \bibnamefont
  {Banfi}}, \bibinfo {author} {\bibfnamefont {Mrinal}\ \bibnamefont
  {Dasgupta}}, \ and\ \bibinfo {author} {\bibfnamefont {Rosa~Maria}\
  \bibnamefont {Duran~Delgado}},\ }\bibfield  {title} {\enquote {\bibinfo
  {title} {{The $a_t$ distribution of the Z boson at hadron colliders}},}\
  }\href {\doibase 10.1088/1126-6708/2009/12/022} {\bibfield  {journal}
  {\bibinfo  {journal} {JHEP}\ }\textbf {\bibinfo {volume} {12}},\ \bibinfo
  {pages} {022} (\bibinfo {year} {2009})},\ \Eprint
  {http://arxiv.org/abs/0909.5327} {arXiv:0909.5327 [hep-ph]} \BibitemShut
  {NoStop}%
\bibitem [{\citenamefont {Banfi}\ \emph {et~al.}(2011)\citenamefont {Banfi},
  \citenamefont {Dasgupta},\ and\ \citenamefont {Marzani}}]{Banfi:2011dx}%
  \BibitemOpen
  \bibfield  {author} {\bibinfo {author} {\bibfnamefont {Andrea}\ \bibnamefont
  {Banfi}}, \bibinfo {author} {\bibfnamefont {Mrinal}\ \bibnamefont
  {Dasgupta}}, \ and\ \bibinfo {author} {\bibfnamefont {Simone}\ \bibnamefont
  {Marzani}},\ }\bibfield  {title} {\enquote {\bibinfo {title} {{QCD
  predictions for new variables to study dilepton transverse momenta at hadron
  colliders}},}\ }\href {\doibase 10.1016/j.physletb.2011.05.028} {\bibfield
  {journal} {\bibinfo  {journal} {Phys. Lett.}\ }\textbf {\bibinfo {volume}
  {B701}},\ \bibinfo {pages} {75--81} (\bibinfo {year} {2011})},\ \Eprint
  {http://arxiv.org/abs/1102.3594} {arXiv:1102.3594 [hep-ph]} \BibitemShut
  {NoStop}%
\end{thebibliography}%

\end{document}